\documentclass[preprint2]{aastex}

\slugcomment{Not to appear in Nonlearned J., 45.}

\shorttitle{CO full depletion}
\shortauthors{Bovino et al.}

\usepackage{natbib}
\usepackage{array}
\usepackage{booktabs}
\usepackage{url}
\usepackage{lipsum}

\begin{document}



\title{The 3D structure of CO depletion in high-mass prestellar regions}




\author{S. Bovino\altaffilmark{1}, S. Ferrada-Chamorro\altaffilmark{1}, A. Lupi\altaffilmark{2}, G. Sabatini\altaffilmark{1,3,4}, A. Giannetti\altaffilmark{4}, and D.~R.~G.~Schleicher\altaffilmark{1}}
\affil{$^1$Departamento de Astronom\'ia, Facultad Ciencias F\'isicas y Matem\'aticas, Universidad de Concepci\'on, Av. Esteban Iturra s/n Barrio Universitario, Casilla 160, Concepci\'on, Chile}
\affil{$^2$Scuola Normale Superiore, Piazza dei Cavalieri 7, Pisa, IT-56126 Italy}
\affil{$^3$Dipartimento di Fisica e Astronomia, Universit\'a degli Studi di Bologna, via Gobetti 93/2, I-40129 Bologna, Italy}
\affil{$^4$INAF - Istituto di Radioastronomia - Italian node of the ALMA Regional Centre (ARC), via Gobetti 101, I-40129 Bologna, Italy}

\email{stefanobovino@astro-udec.cl}


\begin{abstract}
Disentangling the different stages of the star-formation process, in particular in the high-mass regime, is a challenge in astrophysics. Chemical clocks could help alleviating this problem, but their evolution strongly depends on many parameters, leading to degeneracy in the interpretation of the observational data. One of these uncertainties is the degree of CO depletion. We present here the first self-consistent magneto-hydrodynamic simulations of high-mass star-forming regions at different scales, fully coupled with a non-equilibrium chemical network, which includes C-N-O bearing molecules. Depletion and desorption processes are treated time-dependently. The results show that full CO-depletion (i.e. all gas-phase CO frozen-out on the surface of dust grains), can be reached very quickly, in one third or even smaller fractions of the free-fall time, whether the collapse proceeds on slow or fast timescales. This leads to a high level of deuteration in a short time both for typical tracers like N$_2$H$^+$, as well as for the main ion H$_3^+$, the latter being in general larger and more extended. N$_2$ depletion is slightly less efficient, and no direct effects on \mbox{N-bearing} molecules and deuterium fractionation are observed.  We show that CO depletion is not the only driver of deuteration, and that there is a strong impact on $D_{frac}$ when changing the grain-size.
We finally apply a two-dimensional gaussian Point Spread Function to our results to mimic observations with single-dish and interferometers. Our findings suggest that the low-values observed in high-mass star-forming clumps are in reality masking a full-depletion stage in the inner 0.1~pc region. 

\end{abstract}


\section{Introduction}
The early stages of high-mass star-forming regions are characterized by large column densities (10$^{23}$-10$^{25}$ cm$^{-2}$) and temperatures of $T < 20$ K \citep{Rathborne2006,Bergin2007}. These conditions are ideal to favor chemical processes like adsorption of heavy-species on the surface of dust grains, a process also known as depletion or freeze-out \citep[e.g.][]{Caselli1999,Bacmann,Hernandez2012,Fontani2012}.  
While it is clear that polar molecules like H$_2$O together with abundant species like CO go through depletion efficiently, for other molecules, in particular N-bearing species, there is no observational evidence of efficient depletion. A recent paper by \citet{Redaelli2019} points toward evidence of partial N$_2$D$^+$ depletion in L1544, but also discusses the limited spatial resolution of their data.
One of the most relevant chemical process which is boosted by the removal of CO from the gas-phase, is the deuterium fractionation ($D_{frac}$), i.e. the enrichment of deuterated molecules relative to their non-deuterated counterpart \citep[see e.g.][]{Ceccarelli2014}. The H$_3^+$ deuterated forms represent the most abundant ions in these regions, and are considered the best tracers of the early stages of the star-formation process \citep{Pagani1992,Walmsley2004,Harju2006,Caselli2008,Giannetti2019}. Estimating the timescales for CO depletion is then relevant to understand how long the early stages last compared to the low-mass counterparts, and how reliable the deuteration is as a chemical clock.

However, measuring the depletion of a molecule is not easy, mainly due to the limited resolution in the observations and line-of-sight contamination. 
In fact, emission from CO when highly depleted is weak, and then line-of-sight integration can only give us a rough estimate of the CO presence in an observed source. Alternative methods to retrieve radial profiles of CO based on local quantities like volume densities have been proposed \citep[see][]{Pagani2012}, but these methods are strongly dependent on the assumed theoretical model and difficult to generalize.

Due to the larger amount of dense gas, in general, depletion is more effective in high-mass star-forming regions than in their low-mass counterparts. An estimate of the CO depletion is usually obtained through the so-called depletion factor, $f_\mathrm{dep}$, that is the ratio between a given CO canonical abundance, and the observed CO \citep{Fontani2012,Giannetti2014}. Values of $f_\mathrm{dep}$ in between 2-80 have been reported in different studies \citep{Fontani2012,Giannetti2014,Feng2016,Sabatini2019}, but these values are generally affected by several assumptions related for instance to the dust opacity and the dust-to-gas ratio \citep[see][for a detailed discussion]{Giannetti2014}. Larger values ($f_\mathrm{dep} > 100$) have been reported by \citet{Zhang2009} in their high-resolution ($\sim 10^{-2}$ pc) observations. 


CO-depletion has been treated with different flavours in theoretical studies. 
\citet{Kong2015} explored  both a constant depletion factor as well as a proper depletion treatment in constant density models similar to \citet{Sipila2015}, concluding that the results are well comparable. However, freeze-out is strongly density-dependent, and constant density models cannot fully account for its evolution in a dynamical environment, with strong consequences also on the deuteration timescale. 

The only existing three-dimensional numerical simulations have either employed approximate chemical models \citep{Goodson} or assumed instead full depletion, i.e. 100 \% of the CO frozen out on the surface of grains \citep{Koertgen2017,Koertgen2018}. 
In the former case, timescales can be overestimated as the chemistry is not evolved alongside the dynamics, ignoring non-linear behavior of the density structure, while in the latter they might be shorter because a full depletion stage tends to favor deuteration.
However, \citet{Sabatini2019}, by employing APEX observations and a toy model, showed that the radius of full depletion in their sources varies in between 0.02 and 0.15~pc, well comparable with the core size in \citet{Koertgen2017} simulations. 
\citet{Ford2011} have pursued a similar study as \citet{Sabatini2019},  for a sample of low-mass star-forming regions, and reported values of depletion in between 4 and 1000, and a depletion radius of 0.02-0.05~pc. However, the same authors state that choices of temperature profile and canonical abundance can strongly affect the results and that they are not able to distinguish between a depletion factor of 10 and a factor of 1000. On the other hand, their constraints on the depletion radius are more robust.

Overall, the dynamical complexity of the star-formation problem (magnetic fields, turbulence, radiation), and the large chain of chemical reactions that should be taken into account  (isomers, isotopes, gas-grains interactions), have led to a mix of non-conclusive results, in particular for what concerns the deuteration timescale, and its possible role as a chemical clock.
In this work, we aim to provide a clearer picture where the complex dynamics along with a detailed model of chemistry is taken into account.

This paper is organised as follows: we briefly introduce the numerical setup for our simulations and sketch the main chemical processes employed, then discuss the main results going from small scales (at core level) to larger scales (clumps), and finally compare the latter with observations.

\begin{table}[t]
   \centering
   \caption{Fiducial values of important physical parameters.}
   \label{tab:ICs1}
   \begin{tabular}{ll}
   \hline Parameter & Value \\\hline
   Temperature$^{a}$ $T$ & 15 K \\
   CR ionization rate $\zeta_\mathrm{CR}$ & $2.5\times10^{-17}$ s$^{-1}$\\
   Dust grain density $\rho_0$ & 3 g cm$^{-3}$ \\
   Dust-to-gas mass ratio $\mathcal{D}$ & $7.09\times10^{-3}$ \\
   Mean molecular weight $\mu$ & 2.4\\\hline%
   \end{tabular}
   \\\flushleft $^a$ We assume $T_{\rm gas} = T_{\rm dust}$ throughout this work.
\end{table}

\begin{table}[t]
   \centering
   \caption{Employed binding energies for the selected species that go through time-dependent depletion (see Wakelam et al. 2017).}
   \label{tab:E_B}
   \begin{tabular}{cc}
   \hline Species & $E_{\rm{B}}$ [K] \\\hline
   N & 720 \\
   O & 1600 \\
   CO & 1300 \\
   N$_2$ & 1100 \\\hline
   \end{tabular}
\end{table}

\section{Methods and numerical setup}
In these simulations, we employ \textsc{gizmo} \citep{Hopkins2015}, 
by using its mesh-less finite-mass method, and the standard cubic-spline kernel, setting the desired number of neighbours to 32. Gravity is based on a Barnes-Hut tree, as in  \textsc{Gadget3} and  \textsc{Gadget2} \citep{Springel2005}, and the non-equilibrium chemistry is solved via \textsc{krome} \citep{Grassi2014}, which has been coupled to \textsc{gizmo} and already employed in a series of papers \citep[e.g.][]{Lupi2018,Lupi2019}. 

Our state-of-the-art network includes 134 species and 4616 reactions. 
This is the largest deuteration network ever employed in three-dimensional simulations.
Compared to the basic network employed in our previous works 
\citep{Koertgen2017,Koertgen2018}, we have now included all the recent updates \citep{Hugo2009,Pagani2009,Sipila2015}.
We follow the evolution of the most relevant C-N-O bearing species like N$_2$H$^+$, CO, HCO$^+$ and their deuterated forms. 
Due to the exponential growth of the network we decided to include molecules up to three atoms only, therefore complex molecules like methanol or ammonia are not included\footnote{We have benchmarked our network towards \citet{Kong2015} and \citet{Sipila2010} obtaining very good agreement with the largest differences within a factor of 2.}.
Depletion of CO, N$_2$, N, and O is treated time-dependently with the standard formula by \citet{Hasegawa1992}.

\begin{table}[t]
   \centering
   \caption{Fiducial initial elemental abundances with respect to the abundance of atomic hydrogen $n_{\rm H}$. The H$_2$ ortho-to-para ratio is set to 3.}
   \label{tab:ICs2}
   \begin{tabular}{cc}
   \hline Species & $n_{\rm species}/n_{\rm{H}}$ \\\hline
   o-H$_2$ & $3.75\times10^{-01}$ \\
   p-H$_2$ & $1.25\times10^{-01}$ \\
   HD & $1.60\times10^{-05}$ \\
   He & $1.00\times10^{-01}$ \\
   o-H$_3^+$ & $1.80\times10^{-10}$ \\
   p-H$_3^+$ & $3.00\times10^{-09}$ \\
   N & $1.05\times10^{-05}$ \\
   O & $1.36\times10^{-04}$ \\
   CO & $1.20\times10^{-04}$ \\
   N$_2$ & $5.25\times10^{-06}$ \\
   \hline
   \end{tabular}
\end{table}

\begin{table*}
    \centering
    \caption{Dynamical parameters employed in the performed runs: Bonnor-Ebert Mass ($M_{BE}$), radius ($R_{BE}$), virial parameter $(\alpha_{vir}$), Mach number $\mathcal{M}$, average magnetic field $\langle B \rangle$, and Jeans masses contained in the Bonnor-Ebert sphere $M_J$. In addition, we report the peak central density $n_0$, the average density calculated from the total mass over the total volume, $\langle n \rangle$, and the free-fall time obtained on the basis of the latter. We note that this is an ideal free-fall time.}\label{table:table1}
    \begin{tabular}{lllccccccc}
    \toprule\toprule
         & $M_{\mathrm{BE}}$ & $R_{\mathrm{BE}}$ & $\alpha_{\rm vir}$ & $\mathcal{M}$ & $\langle B \rangle$ & $M_J$ & $t_{\rm ff}$ & $n_0$ & $\langle n \rangle$  \\
         & [M$_\odot$] & [pc] & & & [$\mu$G] & & [kyr] & [cm$^{-3}$] & [cm$^{-3}$] \\\midrule
        \multicolumn{9}{l}{\textbf{Cores:}}\\
        M0 & 60 & 0.17 & 1.44 & 3 & 137 & 27 & 150 & $1.16 \times 10^6$ & 4.89 $\times\, 10^4$  \\
        
        
        M1 & 20 & 0.17 & 4.32 & 3 & 46 & 5 & 260 & $1.81 \times 10^5$ &$2.21\times 10^4$ \\
        \multicolumn{9}{l}{}\\
        \multicolumn{9}{l}{\textbf{Clumps:}}\\
        M2 & 160 & 0.7 & 0.64 & 1.6 & 22 & 14 & 767 &$4.34 \times 10^5$ &  $1.88\times 10^3$\\
        M3 & 160 & 0.7 & 24.7 & 10 & 22 & 14 & 767 & $4.34 \times 10^5$ & $1.88\times 10^3$\\
        \bottomrule\bottomrule
    \end{tabular}
\end{table*}

\begin{equation}
	k_\mathrm{ads} (X) = \pi \langle a^2 \rangle  v(X) n_\mathrm{dust}
\end{equation}
with $\langle a^2 \rangle$  being the average over a grain-size distribution\footnote{We assume here a typical MRN grain-size distribution \citep{Mathis1977}, with a power law index of 3.5 and $a_{min} = 1.0\times 10^{-7}$ cm and $a_{max} = 2.5\times 10^{-5}$ cm.}, $v(X) = \sqrt{8 k_\mathrm{B} T_\mathrm{gas} / \pi m_X}$ the thermal speed of the species $X$ with mass $m_X$, and \mbox{$n_\mathrm{dust} = \mathcal{D} n_{gas} m_\mathrm{H}\mu/M_{dust}$} the dust number density, with $n_{gas}$ the gas number density. 
We consider dust grains to be silicates with a typical specific density $\rho_0 = 3$ g cm$^{-3}$, and a dust-to-gas mass ratio of $\mathcal{D}\sim 7.09 \times 10^{-3}$ as reported in \citet{Kong2015}. The gas mean molecular weight is $\mu = 2.4$.  We report in Table \ref{tab:ICs1} the most relevant parameters. In section \ref{sec:parameters} we show how much the results change when we assume a typical average size $\langle a \rangle = 0.1$ $\mu$m, without averaging over a proper distribution.

As we are exploring low temperature regimes ($T = 15$ K) the only relevant contribution to desorption is the cosmic-ray induced desorption, defined as
\begin{equation}
	k_\mathrm{des} (X) = \nu_0 \exp{(-E_D(X) / k_\mathrm{B} T)} f(70\, \mathrm{K})
\end{equation}\label{eq:cr_des}

\noindent where $f(70\, \mathrm{K})$ is the fraction of time the grains spend at 70 K, $\nu_0$ the harmonic oscillator frequency ($\sim$10$^{12}$~s), and $E_D(X)$ the binding energy of the $X$th species. We employ the old $f(70\, \mathrm{K}$) function by \citet{Hasegawa1992} but we notice that new calculations have been recently reported by \citet{Kalvans2016}, pointing out that a lower temperature (i.e. 40-50 K) is more likely to induce desorption, and for longer times\footnote{We will further explore this effect in a future work.}. The binding energies for the main species which go through adsorption/desorption processes are taken from \citet{Wakelam2016} and reported in Table \ref{tab:E_B}.


\subsection{Initial conditions\label{sec:chem}}

We initialise an isothermal ($T = 15$ K) Bonnor-Ebert (BE) sphere by solving the Lane-Emden equation, and  then rescaling the density profile to match the physical properties of the desired core/clump. The BE sphere is defined by the mass ($M_\mathrm{BE}$), the radius ($R_\mathrm{BE}$), and the central density ($n_0$).

Turbulence is initialized following a Burgers-like power spectrum \citep[see][]{Koertgen2017}. After the map has been generated, the velocity field is renormalised 
according to the gas temperature and the desired Mach number~$\mathcal{M}$.

The magnetic field is aligned with the \mbox{$z$-direction} and scales as 

\begin{equation}
B_z(R_\perp) = B_0 (1 + R_\perp/R_\mathrm{BE})^{-\kappa}
\end{equation}

\noindent where $\kappa$ = 1.5 is the scaling exponent and $B_0$ is the value of the magnetic field at $R_\perp=0$ \citep[see also][]{Koertgen2017}, obtained by enforcing a user-defined mass-to-flux ratio \mbox{$\mu=N_\mu \mu_{crit}$}, where $\mu_{crit}$ is the critical mass-to-flux ratio defined as \mbox{$\mu_{crit} = \sqrt{G}M_\mathrm{BE} / 0.13\pi R_\mathrm{BE}^2 \langle B \rangle$}. 

We assume fully molecular initial conditions for the chemical species as reported  in Table \ref{tab:ICs2}. Our initial H$_2$ ortho-to-para ratio (OPR) is 3, which is a very conservative assumption as at the densities considered in the simulations the OPR should be around 0.1-0.01 \citep[see e.g.][]{Trompscot2009}.
The cosmic ray ionisation rate (CRIR) is set to $\zeta_{\mathrm {CR}} = 2.5 \times 10^{-17}$  \citep{Kong2015}.

N$_2$ was initialised to contain half of the total nitrogen abundance, while half remains in atomic form. We tested a fully molecular nitrogen case, and the results do not change too much, as also shown by \citet{Kong2015}.

We explore the effects of the cosmic ray ionisation rate (CRIR) and the initial H$_2$ OPR on depletion and deuteration in Section \ref{sec:parameters}.

\begin{figure*}[ht]
	\centering
	\includegraphics[scale=0.7]{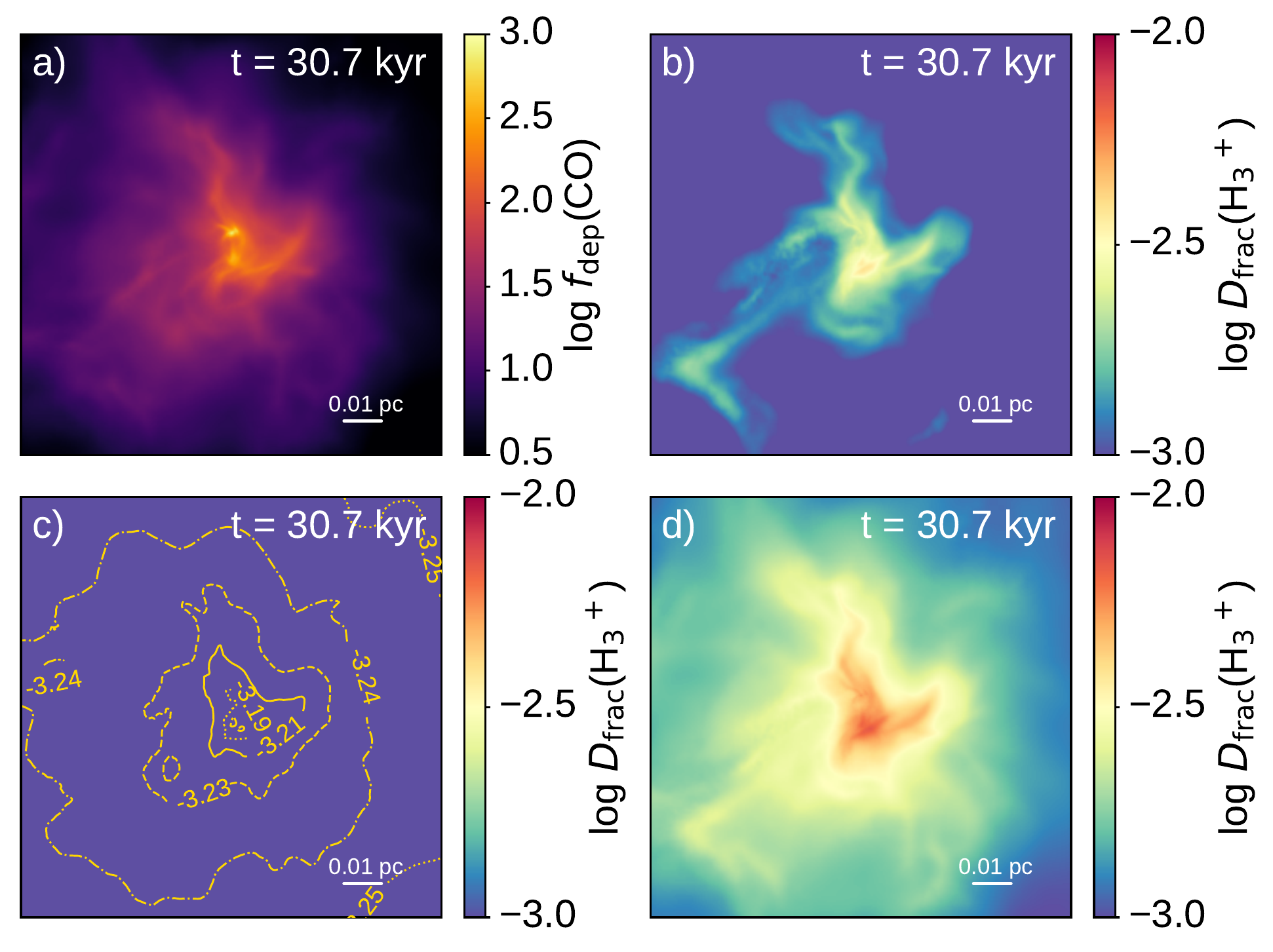}	
	\caption{Top: Column density-weighted quantities for the reference core: depletion factor (a), and deuterium fractionation of H$_3^+$ (b) for the reference case where the CO depletion is followed in time. Bottom: deuterium fractionation of H$_3^+$ (c) by assuming a fixed depletion factor of $f_\mathrm{dep} = 10$, and the case where depletion has not been included (d), i.e. a fully depleted case. The column densities out of which the reported quantities have been calculated are obtained from an integration over a 0.2 parsec line-of-sight.}\label{fig:dep_analysis}
\end{figure*}

\section{Results}
In the following subsections we will present our main results
encompassing different mass ranges, turbulent Mach numbers, and spatial scales. The parameters employed in the different runs are reported in Table~\ref{table:table1}. The number of particles is arranged to keep the same mass resolution, i.e. 2$\times 10^{-4}$ M$_\odot$. After defining the mass resolution, the maximum spatial resolution is determined by the softening parameter $\epsilon$, corresponding to the minimum interparticle distance for which the calculated gravitational force is Newtonian\footnote{We employ adaptive softening that ensure hydrodynamics and gravity are solved assuming the same mass distribution within the kernel.}.  By employing the formula reported by  \citet{Hopkins2017}, with $n_\mathrm{sink} = 2.7 \times 10^9$ cm$^{-3}$ (the density threshold above which a sink particle is created), we obtain a spatial resolution of roughly 10$^{-4}$ pc (i.e. $\sim$20~AU). We note that the entire analysis of the results is done before the sink formation to avoid artificial effects on the chemistry.

\begin{figure*}[!ht]
	\centering
	\includegraphics[scale=0.7]{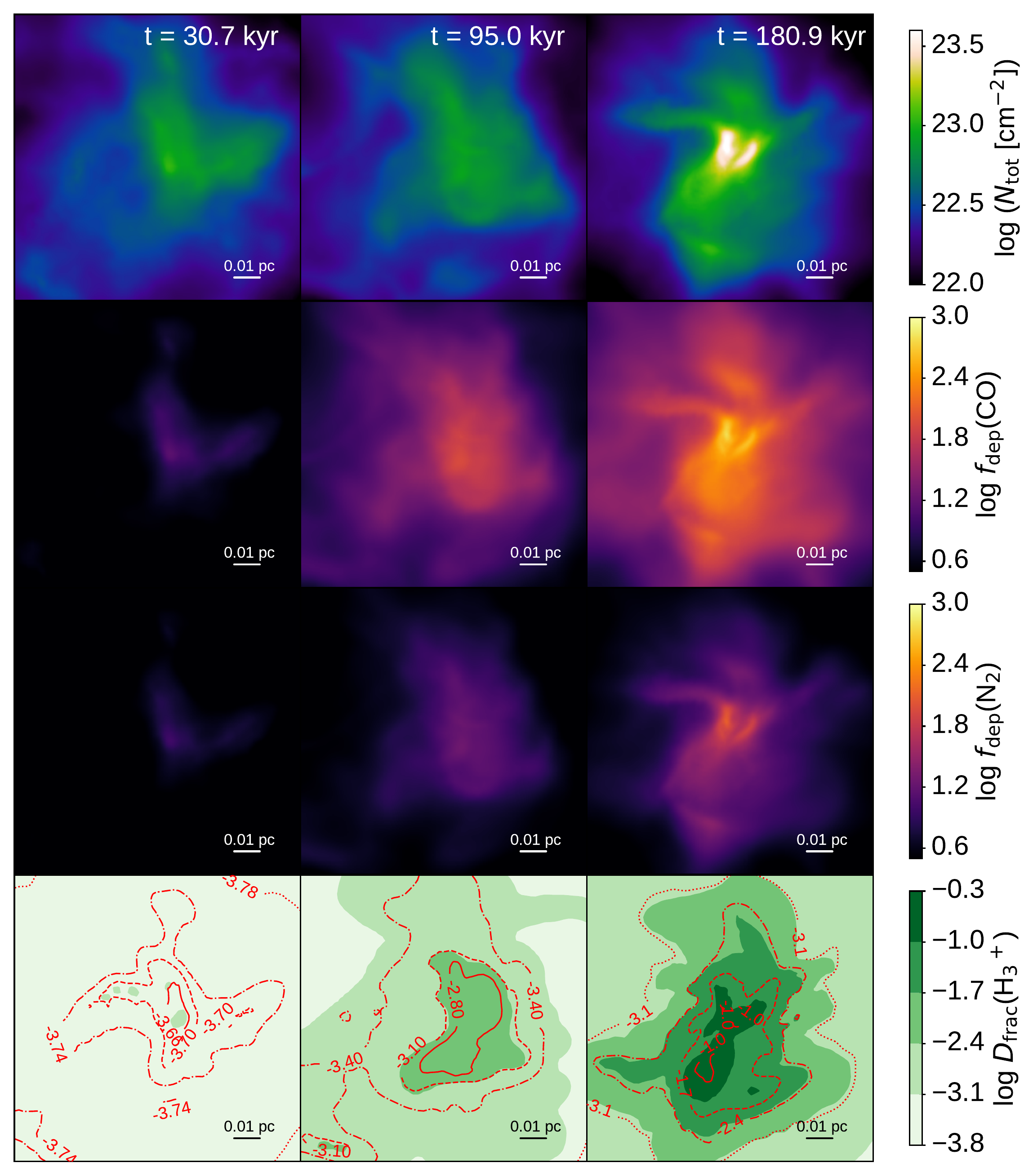}	
	\caption{From top to bottom: total column density, column density-weighted CO and N$_2$ depletion factors, and H$_3^+$ deuterium fractionation with superimposed (red contours) N$_2$H$^+$, at three different evolution times for the 20 M$_\odot$ core.}\label{fig:mass_analysis}
\end{figure*}

\subsection{Effect of depletion treatment}
We start by comparing different treatments of CO depletion for the reference run (M0). This is a 60 M$_\odot$ core, highly unstable but strongly supported by turbulence (slightly supervirial with $\alpha_{vir} = 1.44$) and magnetic pressure (\mbox{$\mu/\mu_{crit}$=2}, and \mbox{$\langle B \rangle = 137 \mu$G}). We report in Fig.~\ref{fig:dep_analysis} the time dependent depletion run (top panels), a case with a constant depletion factor (i.e. \mbox{$f_\mathrm{dep} = 10$}, panel c), quite often employed in simple one-zone models (see e.g. \citealt{Kong2015}), and a full depletion case (panel d) as used in previous works (e.g. \citealt{Koertgen2017}). The quantities are obtained from an integration over a 0.2 parsec line-of-sight (LoS). 

With this set of simulations we are able to disentangle the effects of an accurate treatment of depletion against a simplified approach. The amount of CO frozen-out on the surface of dust grains is very high and reaches values of 1000 in the central part in only 30~kyr.  This leads to  deuterium fractionation values around 0.01 in the central region. When a constant amount of CO is considered to be frozen out (in this case 90\%, i.e. $f_\mathrm{dep} = 10$), the deuterium fractionation of H$_3^+$ drastically changes, with the values much smaller sticking around 5$\times$10$^{-4}$ (panel c). This means that the assumption of a constant depletion factor can strongly affect the models and the interpretation of observational data, overestimating the amount of CO left in gas-phase, which normally will continue to be adsorbed over time, while the collapse proceeds. We finally report in panel-d the case where full depletion has been assumed, i.e. no CO in gas phase ($f_\mathrm{dep} = \infty$). This test suggests us that a full depletion assumption at small scales is reasonable, as the deuterium fractionation is not so different from the case where we accurately follow the CO depletion in time. With our results we prove that the depletion process, strongly density-dependent, is indeed very fast at small scales where densities easily reach values of about 10$^8$ cm$^{-3}$, making the assumption of full depletion good enough to be used in very expensive three-dimensional simulations. The deuteration fraction between the time-dependent freeze-out case and the fully depleted case is very similar in the central 0.01 pc and changes by maximum 40\% on larger scales. This result is also supported by recent observational work by \citet{Sabatini2019}, which showed that full depletion conditions can be reached for scales below $\sim$0.15~pc, not different from those we report in these simulations.

\begin{figure}[!ht]
	\centering
	\includegraphics[scale=0.465]{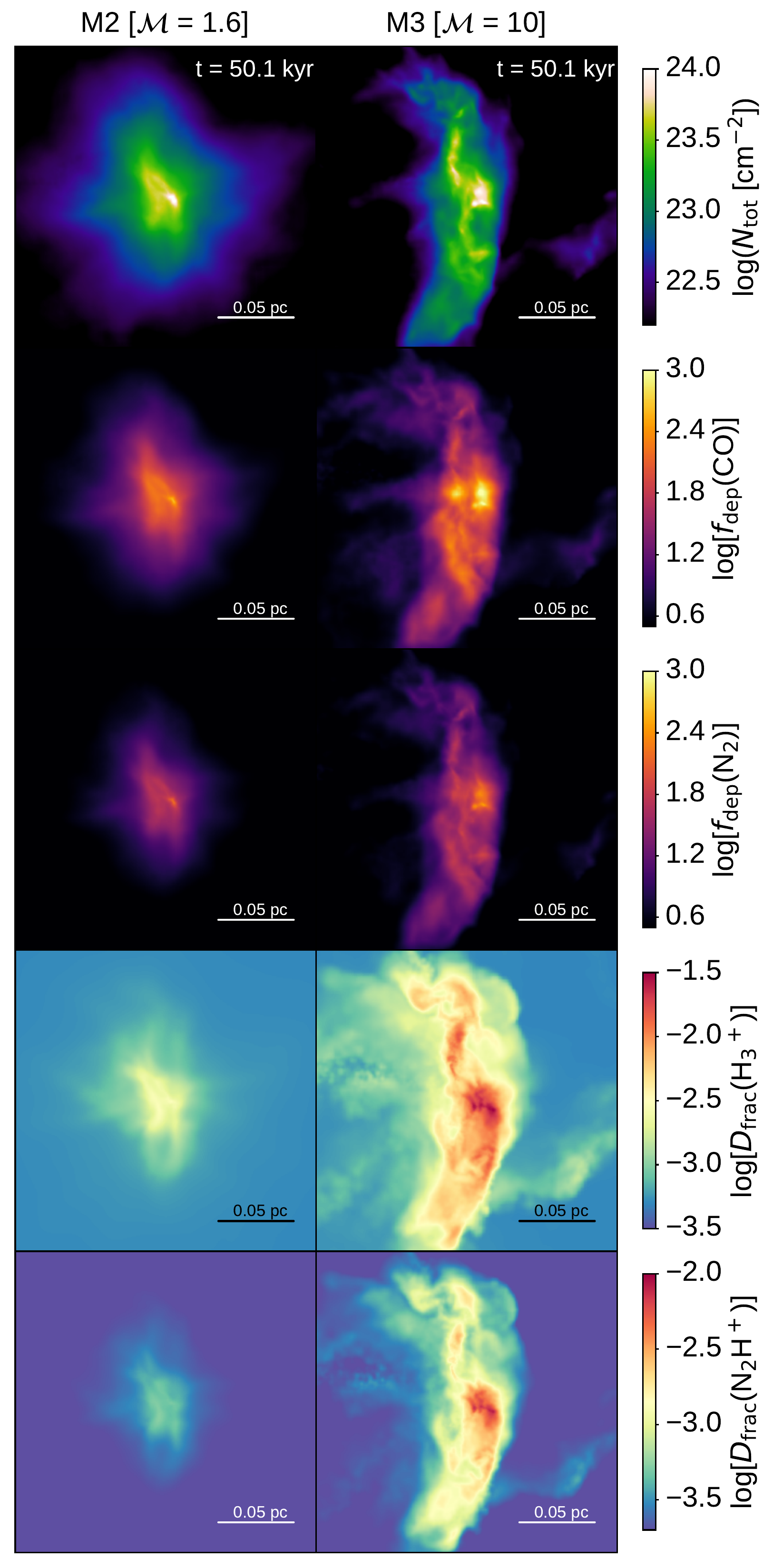}\\
	\caption{From top to bottom: total column density, column density-weighted CO (map) and N$_2$ (white contours) depletion factor, H$_3^+$ deuterium fractionation, and N$_2$H$^+$ deuterium fractionation at t = 50 kyr, for the clump with $\mathcal{M} = 1.6$ (left), and the more turbulent case with $\mathcal{M} = 10$ (right).}\label{fig:turbulence}
\end{figure}

\subsection{Slow collapse at core level}
To explore timescales for the different processes we have initialised a less dense core (M1 in Table~\ref{table:table1}), highly supervirial. The results are reported in Fig. \ref{fig:mass_analysis} at three different times. From top to bottom panels we show the total column density, the depletion factor (for both CO and N$_2$), and the deuteration fraction for H$_3^+$ (map) and N$_2$H$^+$ (red contours), respectively. To reach high values of depletion compared to the core M0 (Fig. \ref{fig:dep_analysis}) more time is needed. At 30 kyr, for instance, the depletion process is slowly started, while after 95 kyr the values of $f_\mathrm{dep}$ are around a hundred on an extended region. In the fast collapse case we already reached a full depletion stage ($f_\mathrm{dep} > 100$) in the inner part of the core, on scales of 0.05 pc after only 30 kyr. The slow collapse core shows larger values of the depletion factor at around 180~kyr on a larger scale. This effect is induced by the fact that the core has more time to become denser also in the outer region. Nevertheless,  in one third of the free-fall time also this highly turbulent core is able to reach high values of deuteration in both tracers, with H$_3^+$ being more extended and larger than N$_2$H$^+$. Typical values of $D_{frac}$ are in between 0.01-0.5 in the entire region. These are in general larger than the fast collapse case, but we should consider that the time evolution we span is in both cases much shorter than typical dynamical timescales. N$_2$ depletion, is slightly slower than CO. This is mainly due to the time lag needed to form N$_2$ via slow neutral-neutral reactions \citep{Nguyen2018}, which allows to have a continuous formation of N$_2$, compared to CO, which forms quickly but also depletes faster \citep[see also][]{Pagani2012}.



 \begin{figure*}[!ht]
	\centering
	\includegraphics[scale=0.5]{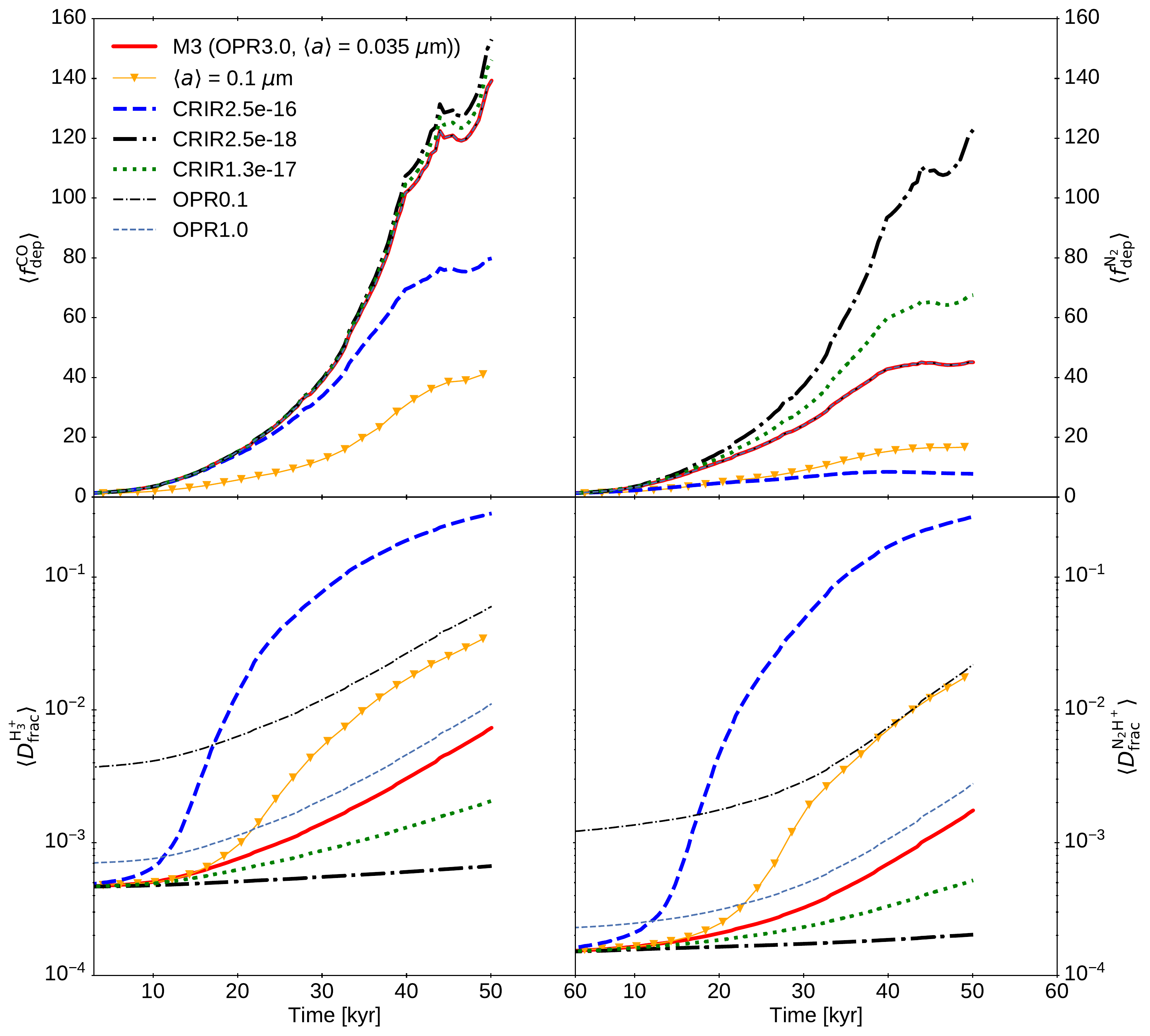}
	\caption{Time evolution of column density-weighted average for the CO and N$_2$ depletion factor, $\langle f^{\mathrm{CO}}_{dep}\rangle$ (on the top left) and $\langle f^{\mathrm{N_2}}_{dep}\rangle$ (top right panel), respectively, H$_3^+$ (bottom left panel) and N$_2$H$^+$ (bottom right panel) deuterium fractionation for different parameters: H$_2$ ortho-to-para ratio, cosmic-ray ionisation rate, and grain-size. The comparison between different CRIR and OPR runs are done with our reference grain-size $\langle a \rangle = 0.035 \,\, \mu$m. The only run with different grain-size is clearly reported in the legend.}\label{fig:parameters}
\end{figure*}

\subsection{Clumps and the effect of turbulence}

To see how the depletion timescale and the collapse change on different spatial scales we performed simulations of a collapsing massive clump with $M_{BE} = 160$ M$_\odot$, and a radius of 0.7 pc (M2 in Table \ref{table:table1}). We kept the mass-to-flux ratio equal to the core simulations to allow for the same magnetic support. In this clump we are subvirial and we expect the collapse to proceed fast.

In Fig. \ref{fig:turbulence} left column, we plot maps of the total column density, the column density-weighted CO and N$_2$ depletion factors, and the deuterium fractionation (both in H$_3^+$ and in N$_2$H$^+$). The collapse proceeds monolithically, as expected for fast collapse conditions, and the depletion reaches high values in roughly 50 kyr (just six percent of the free-fall time), but on very small scales ($<$ 0.05 pc). The deuteration is fast and reaches values larger than 10$^{-3}$ in H$_3^+$, while N$_2$H$^+$ similarly to the core case proceeds slower. We have to consider, once again, that dynamically this clump is very young, as it evolved only for six percent of the free-fall time. We expect deuteration to increase quickly in a short time.

The reference clump (M2) has a Mach number of 1.6 and it is subvirial. 
We ran a second clump (M3) with $\mathcal{M} = 10$, making the clump highly supervirial ($\alpha_\mathrm{vir} \sim 24$) to explore the effect of turbulence. In Fig.~\ref{fig:turbulence} right column, we report the results at 50~kyr, and compare them to the reference massive clump (M2). We first note that the clump slowly collapses in a filamentary fragmented structure as seen from the total column density map. Due to the ongoing fragmentation, the column density-weighted depletion factor is larger and more extended compared to the monolithic collapse case (M2). In this specific case also the depletion factor for N$_2$ reaches high values in the inner part. In both clumps the CO depletion factor is larger than 100, i.e. more than 99.9\% of the CO is frozen-out on the surface of dust grains, but on different scales. In the same figure we also show the deuterium fractionation of N$_2$H$^+$ and H$_3^+$. While both tracers show high level of deuteration, close to observed values, H$_3^+$  is in general more widely distributed. This is in part due to the fact that to form N$_2$D$^+$ a large amount of H$_2$D$^+$ and N$_2$ are needed. 

Overall, the turbulence has a dual effect here, first to slow-down the collapse, but most importantly to enhance fragmentation, with the net result of having larger column densities at core scales and larger depletion and deuteration values compared to a monolithic collapse. We note that filamentary structures have been also observed for example by \citet{Fontani2016}. The fragmentation into low-mass cores seems to point towards global collapse theories with formation of low-mass stars and subsequent accretion to form high-mass protostars \citep[e.g.][]{Motte2017}. However, at this stage, it would be very speculative to rule out other theories even though our simulations go in the direction of a global collapse with turbulence playing a crucial role in the fragmentation of the gas.


\subsection{Grain-size effects and other parameters}\label{sec:parameters}
The grain-size plays an important role in the chemistry of the deuteration because it affects the depletion rate which in turn can delay or speed-up the deuteration process. For the adsorption, in general, $k_{ads} \propto \langle a \rangle^{-1}$, which means that larger  grain radii decrease the adsorption rate and inhibit depletion. On the contrary smaller radii boost the depletion process. This has been already shown by \citet{Sipila2010} in a simple one-dimensional study, and here we confirm the same trend. When employing a size properly calculated from a typical MRN grain-size distribution, i.e. $\langle a \rangle~=~0.035~\mu$m, the depletion values are larger than 100, implying a fully depleted situation. By assuming a typical grain size of 0.1 $\mu$m instead the depletion is reduced by more than a factor of three. This is reported in  Fig. \ref{fig:parameters} (top left), where we show the time evolution obtained from an average over a scale of 20,000 AU. A similar effect is seen also for the N$_2$ depletion factor (on the top right panel). In addition, for N$_2$ we also note a stronger impact of the CRIR, due to the lower binding energy compared to CO. In the same figures (bottom left and bottom right panels) we report the effect of different parameters on deuterium fractionation. The bottom left panel shows the deuterium fractionation for H$_3^+$ and the bottom right panel the one for N$_2$H$^+$. From this figure we see how an increase of  the grain-size  to typical values of 0.1 $\mu$m causes an increase of the deuteration even if depletion decreased. The deuteration increase is not easy to disentangle as there is a mix of processes which can cause this trend. The increase in $D_{frac} (\mathrm{N_2H^+})$ is due to a combined effect: i) a less efficient N$_2$ depletion which implies more nitrogen available in the gas-phase to form N$_2$D$^+$ via N$_2$~+~H$_2$D$^+$ reaction; ii) more atomic deuterium available in gas-phase which boosts the abundance of D$_2$ with an effect on the o-H$_2$D$^+$ abundance via H$_3^+$~+~D$_2$ reaction. In addition, there is also more atomic deuterium which boosts N$_2$D$^+$ via a second reaction N$_2$H$^+$~+~D. 

We show in the same figure also the effect of varying the cosmic-ray ionisation rate (CRIR) and the H$_2$ ortho-to-para ratio (OPR). As expected, and similarly to what was already reported in \citet{Koertgen2018} an increase of the CRIR boosts the deuteration while it has a negative effect on the depletion produced by a faster desorption induced by the cosmic rays. Once again there is an anticorrelation among different parameters which are important for the deuteration. A decrease of the H$_2$ OPR increases the deuteration by almost one order of magnitude when we move from the statistical ratio of three down to typical expected values of about 0.1. No effects on the depletion are reported.

Overall, this analysis suggests us that the CO depletion is not the only driver of deuteration, the OPR and in particular the grain-size can boost deuteration even in the cases with less efficient depletion. 

\begin{figure*}[!ht]
\centering
\includegraphics[scale=0.35]{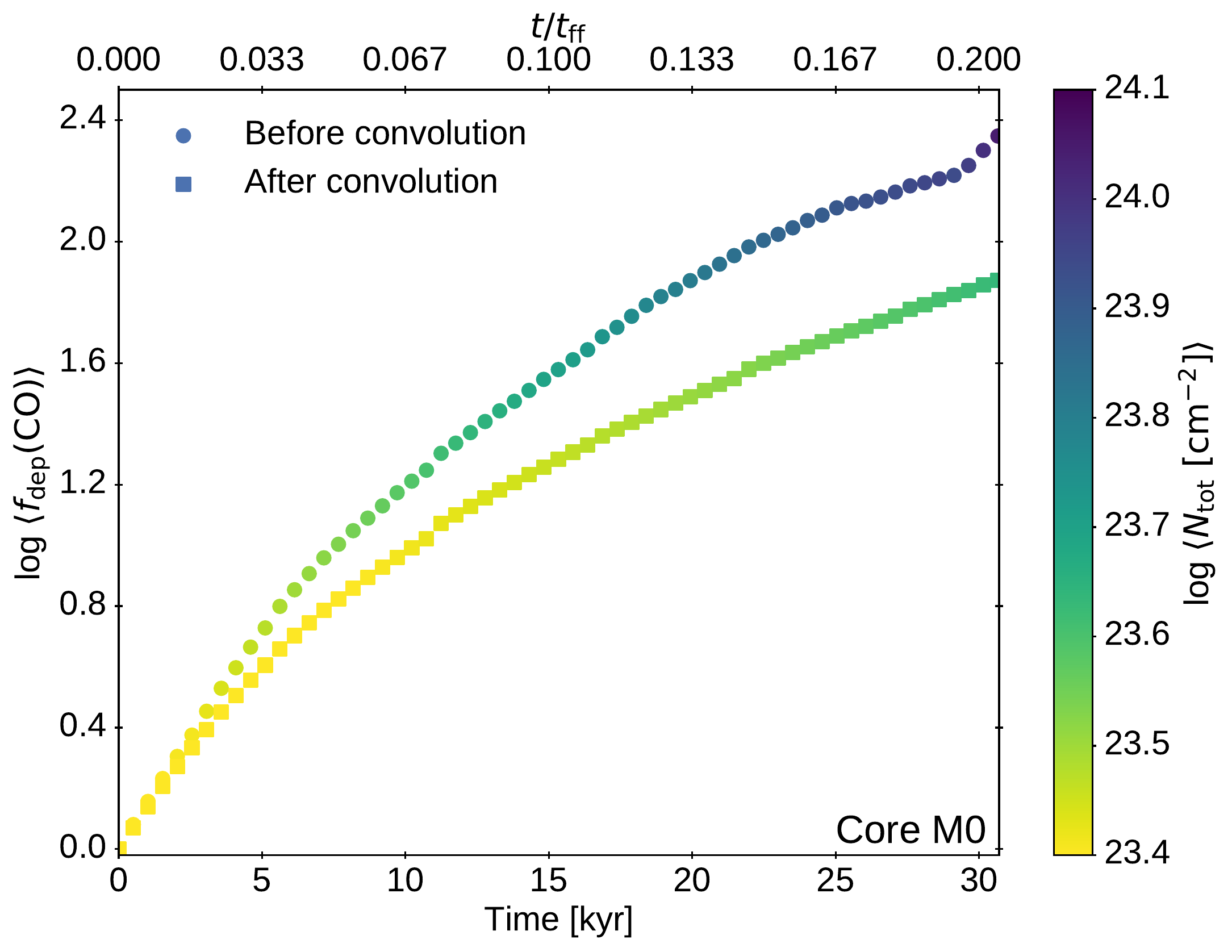}
\includegraphics[scale=0.35]{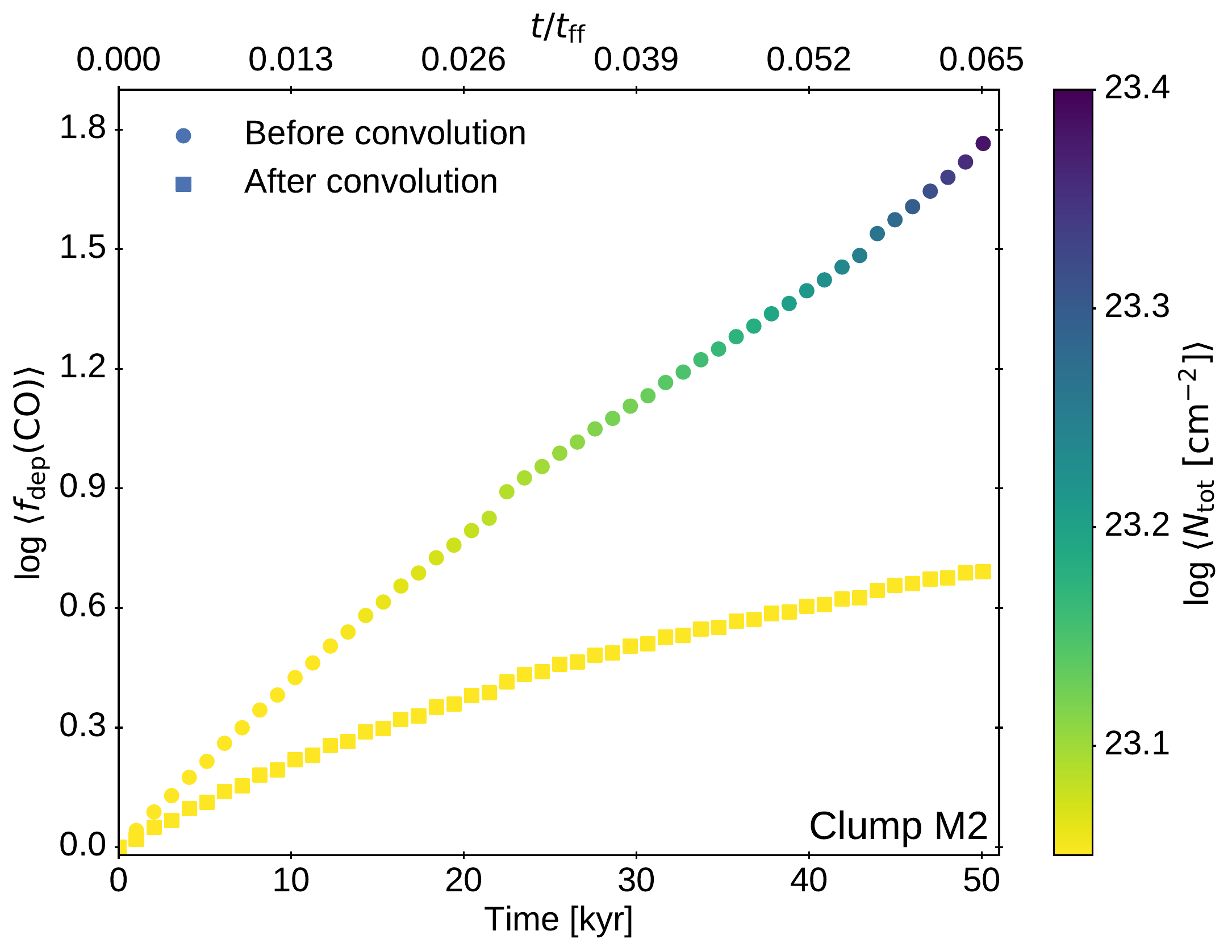}\\
\includegraphics[scale=0.35]{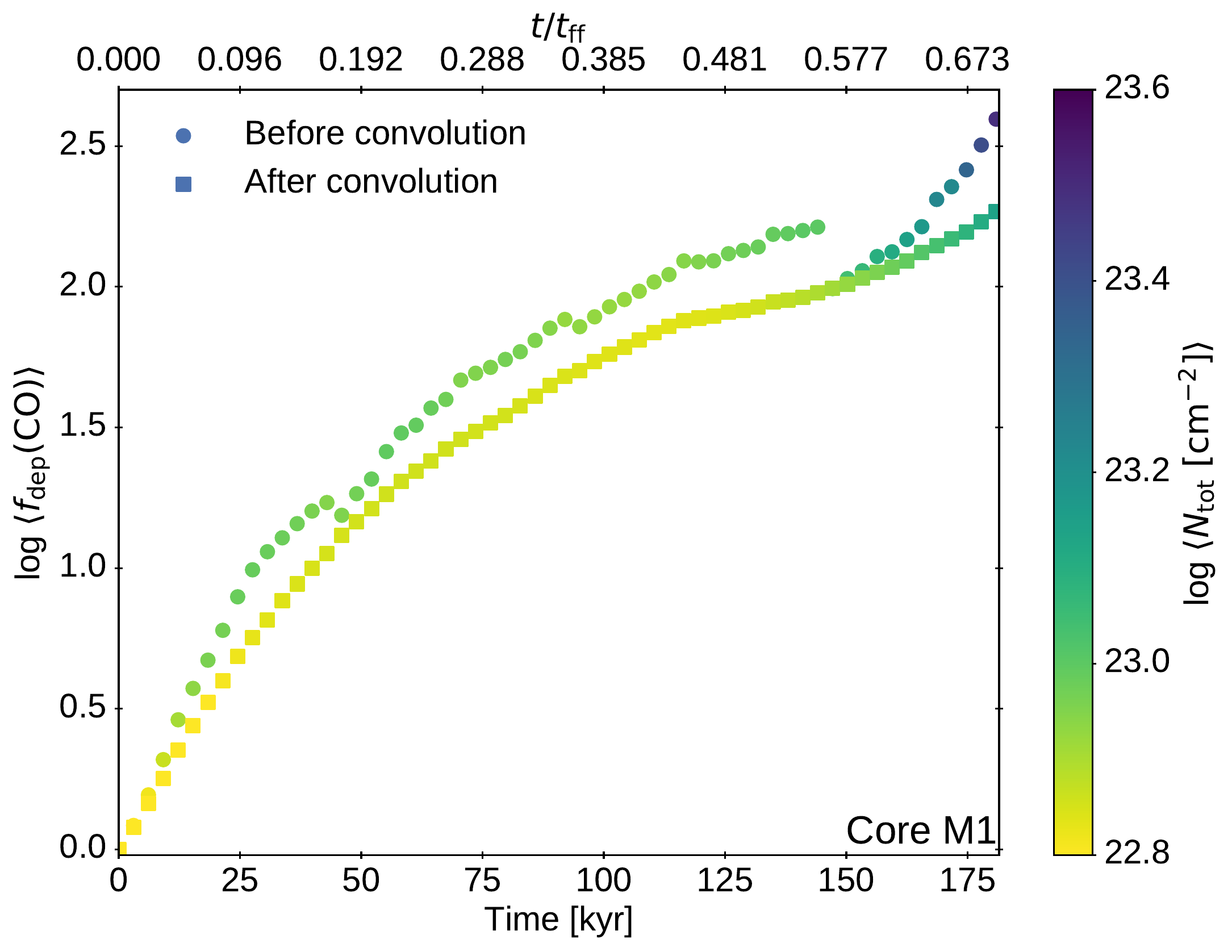}
\includegraphics[scale=0.35]{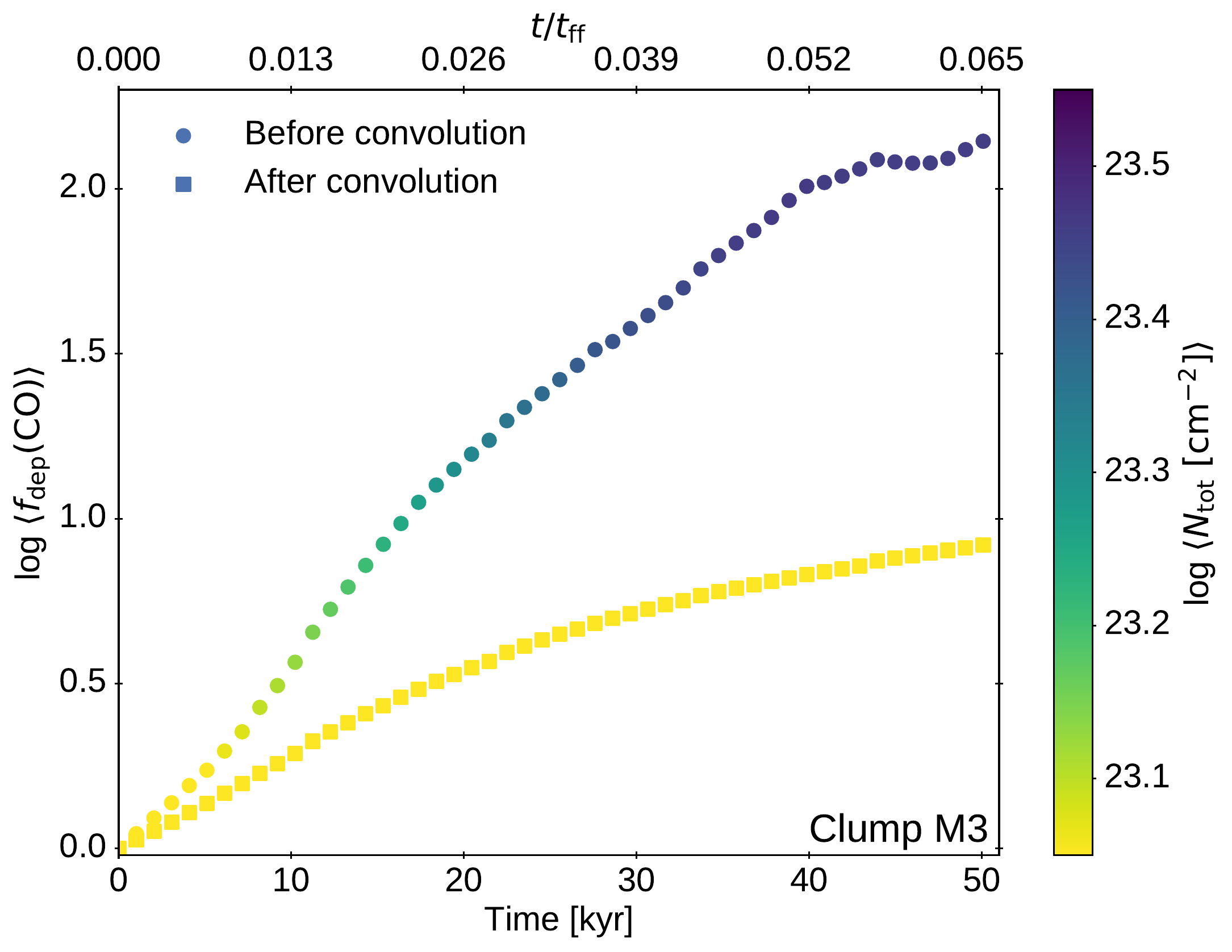}
\caption{Time evolution of the  column density-weighted  CO depletion factor $f_\mathrm{dep}$ for different realizations. Top left: the reference case core with $M = 60$ M$_\odot$, top right: standard clump (M2), bottom left: the slow collapse core with $M=20$ $M_\odot$, and bottom right: the clump with high turbulence ($\mathcal{M} = 10$). The average is calculated over the beamsize (2,000~AU and 20,000 AU for the core and the clump, respectively) centered at $N_{peak}$. In the same figure we reroprt the time evolution for the convolved maps.}\label{fig:analysis1}
\end{figure*}

\subsection{Comparison with observations}
To give a general overview of our results we report in Fig.~\ref{fig:analysis1} the time evolution of the depletion factor $f_\mathrm{dep}$ for the different realizations (M0, M1, M2, and M3). We average over a beamsize of 2,000 AU and 20,000 AU centered on the column density peak, for the cores and the clumps, respectively. The general trend is very similar for all the cases presented: the depletion increases over time. The timescale to reach these values is however different going from 30-50 kyr for fast collapse or situations where fragmentation occurs, to 100 kyr for slow collapse (and less massive) cases. We note, once again, that an $f_\mathrm{dep} = 100$ or larger means that almost the entire CO in gas phase is now on dust ($\sim$99.9\%), i.e. a full depletion case. This stage is reached within very small fractions of the free-fall time, suggesting that full depletion situations are very likely in high-mass star-forming regions and the low observed values are only due to dilution effects. 

\begin{figure}[!h]
\centering
\includegraphics[scale=0.38]{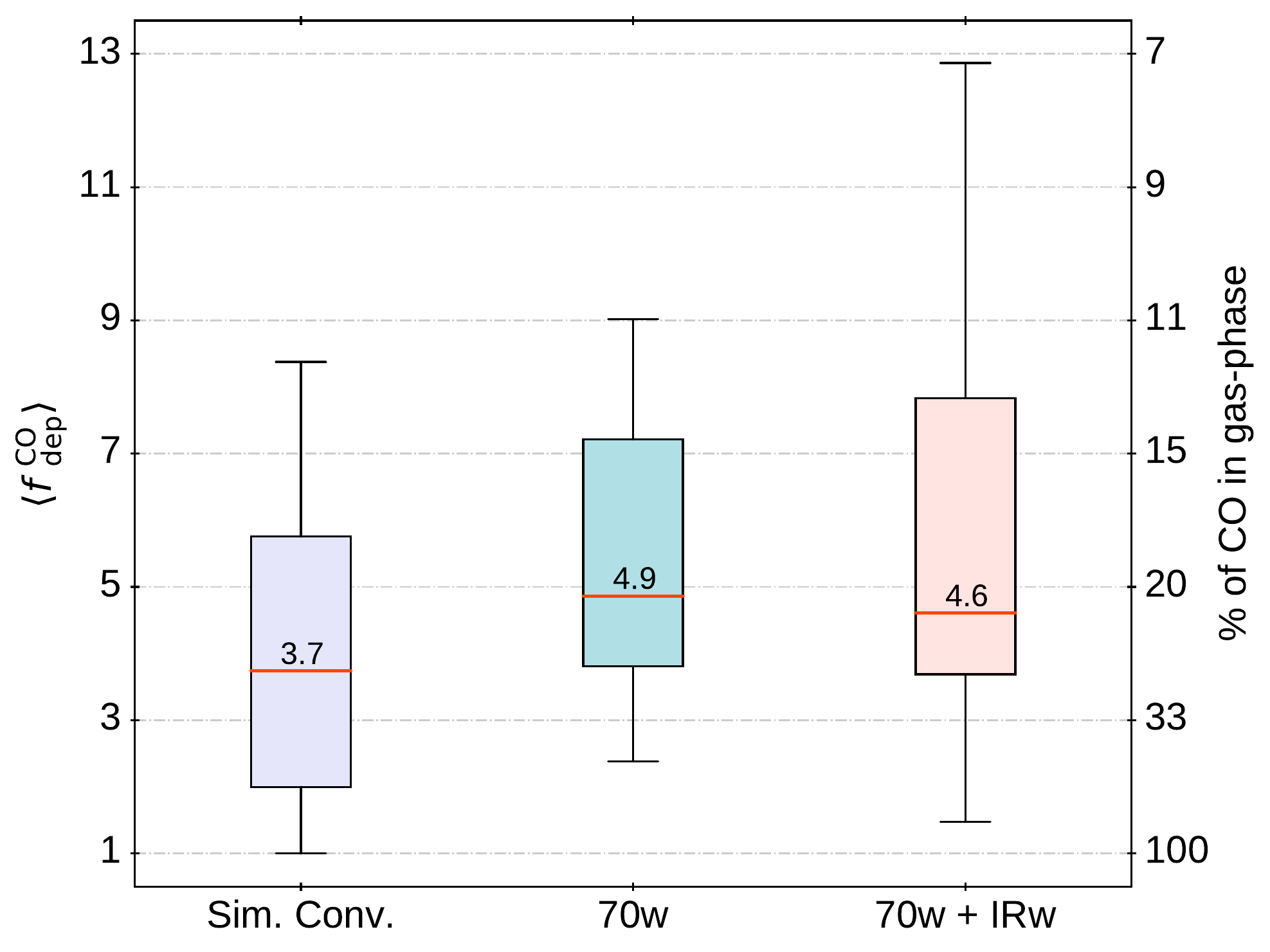}
\caption{Boxplot of the convolved simulations, observations from Top100 \citep{Giannetti2017,Koenig2017} by considering only the 70w sources, and one set by including also the IRw. On the right y-axis we report the percentage of CO which remains in gas phase.}\label{fig:boxplot}
\end{figure}

A comment on the effect of cosmic-rays induced desorption is necessary at this point. This process tends to act against the depletion, releasing CO into the gas-phase. What we see in our simulations is that this effect is very mild due to the high densities reached ($n \sim 10^8$ cm$^{-3}$) in the center. The latter boosts the adsorption process, as its nature is intrinsically collisional and directly proportional to the density. Cosmic-rays induced desorption acts against the depletion but on longer timescales and only when a high-depletion stage has already been reached. Typical timescales for CO cosmic-rays induced desorption can be analytically estimated, and are around 10$^4$~years or longer. In addition, considering that the density reaches high values in a short time, and that cosmic-rays flux should in general be lower in high-density regions (e.g. \citealt{Padovani2009}), we expect this effect to be even weaker.

\paragraph{PSF convolution} To allow for a direct comparison of our depletion values ​​with those from observations, our maps were convolved with a 2D-Gaussian function\footnote{For the convolution the astropy.convolution ``convolve" function was applied, between our maps and the 2D-Gaussian generated by the "Gaussian2Dkernel" function of \textit{astropy.modeling.models}.} considering different FWHM: 2,000 AU, in the case of ALMA-like observations in Band 6, and 20,000 AU in the case of a typical APEX beam. Both FWHMs are calculated assuming the same distance as for G351.77-0.51 (i.e. $\sim$1~kpc, \citealp{Leurini2011}) and the frequency of the \mbox{CO (2-1)} transition. 
The observed depletion factor ​​has been seen to vary between 1 and 6 in maps with a resolution of 36'' \citep{Sabatini2019}. 
We report the results of this convolution in Fig. \ref{fig:analysis1} and compare with the non-convolved ones. Before convolution, depletion factors go up to $f_\mathrm{dep} \sim 10^3$ at a resolution of $\sim 6.6 \times 10^{-5}$ pc/pixel.  Once the convolution is applied, we see that high depletion factors are indeed masked, with changes ranging from a factor of 2 at the core scale (FWHM=2,000 AU) to values which go down by an order of magnitude, leading to typical observed $f_\mathrm{dep}$ of 3-8 (see right panels for the clump cases, FWHM=20,000 AU). This would mean that observed values are actually representing a full depletion stage, which is difficult to observationally retrieve as already shown through different analysis by \citet{Sabatini2019,Ford2011,Pagani2012}. 
We expect that properly solving the radiative transfer equation and produce more realistic synthetic observations will further lower the $f_\mathrm{dep}$ values. This will be explored in a forthcoming paper.

To give a statistical significance to our comparison with observations we take the Top100 sample \citep{Giannetti2017,Koenig2017}, which report CO depletion factors for several high-mass clumps at different evolutionary stages. We select the least evolved, the ones which are dark at 70 and 24 micron, named 70w and IRw,	 respectively and report in Fig. \ref{fig:boxplot} a boxplot, which includes all our simulations for the clumps after convolution. We compare the latter with the 70w only sample (resembling very early stages of the star-formation process), and the mix of 70w and IRw sources, which then include also sources that are more evolved but still at an early stage (with temperatures still below 20 K and luminosities around 100 $L_\odot$). The median of the distribution is very similar both for our theoretical sample and for the observational ensemble, with values around $\langle f^\mathrm{CO}_\mathrm{dep} \rangle = 4$. The minimum and the maximum are quite in agreement, in particular when we consider the 70w only sample, meaning that our simulations are more representative of a very early stage in the star-formation process.

\section{Discussion and Conclusions}
Depletion of heavy elements, in particular CO, has a strong impact on the gas and surface chemistry of prestellar cores. Deuterium fractionation, is one of these processes. In this paper we have reported the first, to date, MHD simulations of high-mass star-forming regions at different scales (clumps and cores) fully coupled with a comprehensive non-equilibrium chemical network to assess the impact of CO depletion timescale on the chemistry, with particular focus on the deuterium fractionation of two important tracers of these regions, H$_3^+$ and N$_2$H$^+$. We have found high levels of CO depletion in all our realizations, with subsequent increase of the $D_{frac}$. N$_2$ depletion turned out to be less efficient, due to its formation channel via slow  neutral-neutral reactions, which competes with the depletion process. N$_2$ evolution is in fact very similar also when we start with full N$_2$ conditions. Values of $f_\mathrm{dep}\mathrm{(CO)}$ in between 50-1000 can be reached in short timescales, fractions of the free-fall time. This would support the theory that a high level of deuteration is built up over a very short time. No specific effects on the timescales of these processes have been observed when going from clumps to cores. On the contrary, the turbulence represents a key physical ingredient in the collapse of these regions. Highly turbulent clumps showed signs of fragmentation, with the newly formed cores reaching high densities. This led to a high level of CO depletion and high deuteration on larger scales, even higher than in the cases of monolithic rapid collapse. An analysis of the effect of different key parameters showed that CO depletion and deuteration are affected by the grain-size. While depletion processes usually decrease when the grain-size is increased, the deuteration has shown a rapid increase. This is mainly due to higher abundances of key species (D, N$_2$, and D$_2$ for example) in gas-phase, which are then available to boost formation channels for H$_2$D$^+$ and N$_2$D$^+$.

Results from a simple PSF convolution would suggest that the low observed values for $f_\mathrm{dep}$ are in reality masking a full depletion stage \citep[see also][]{Pagani2012}.
We have also disentangled the effect of line-of-sight integration and beam-averaged quantities (i.e. before and after convolution), comparing them with the real values coming from the simulations based on local quantities. The loss due to line-of-sight effects (projection) is already very large and we are able to recover only 4\% of the depletion factor at the density peak. When we apply the convolution at different FWHMs (i.e. different telescope resolutions) we recover only a very small portion of the real values, going from 2\% for a FWHMs = 500 AU (typical of ALMA) to 0.02\% for an APEX--like FWHM of 20,000 AU. To appreciate the effect of telescope/spatial resolution, we compare the convolved values with the non-convolved ones taking the column density. In this case, even with a very good resolution (FWHM~=~500~AU) we can recover only 50\% of the depletion factor, while this percentage drops to 0.7\% for FWHM~=~20,000 AU. This is on top of the line-of-sight effects. Overall, the combined effects of integrating over the line-of-sight, and the spatial limits imposed by telescopes resolution, show a dramatic information loss with respect to the real depletion factors. From observations, we can therefore only obtain a lower limit on the amount of depletion, while it seems unlikely that the true values in the central part can be recovered, unless the depleted region is very extended.

To conclude, the observed high-level of CO depletion, together with the very short timescale needed to reach typically observed deuteration ($D_{frac} \sim 0.1$), within the studied parameters space, would question the importance of deuterium fractionation as a reliable chemical clock, unless the early stages of high-mass star-formation are really short. Uncertainties on the initial conditions, and degeneracy on the results should however be kept into consideration and further explored. 


\acknowledgments
\section*{Acknowledgments}
SB is financially supported by CONICYT Fondecyt Iniciaci\'on (project 11170268), CONICYT programa de Astronomia Fondo Quimal 2017 QUIMAL170001, and BASAL Centro de Astrofisica y Tecnologias Afines (CATA) AFB-17002. DRGS thanks for funds through CONICYT PIA ACT172033 and CONICYT Fondecyt regular (project 1161247). SFC thanks Scuola Normale Superiore di Pisa, for the kind hospitality, during his visit in April 2019, where part of this work has been pursued, and CONICYT Programa de Astronom\'ia Fondo ALMA-CONICYT 2017 Project \# 31170002. AL acknowledges support by the European Research Council No. 740120 `INTERSTELLAR'. SB is grateful to Olli Sipil\"a for having shared his chemical network for deuteration studies and to Thomas Millar for fruitful discussions on the chemistry of deuteration. 
The simulations were performed with resources provided by the \textit{KULTRUN Astronomy Hybrid Cluster} at Universidad de Concepci\'on.


\begin{thebibliography}{}
\expandafter\ifx\csname natexlab\endcsname\relax\def\natexlab#1{#1}\fi

\bibitem[Bacmann et al.(2002)]{Bacmann} Bacmann, A., Lefloch, B., Ceccarelli, C., et al.\ 2002, \aap, 389, L6

\bibitem[Bergin, \& Tafalla(2007)]{Bergin2007} Bergin, E.~A., \& Tafalla, M.\ 2007, \araa, 45, 339

\bibitem[Caselli et al.(1999)]{Caselli1999} Caselli, P., Walmsley, C.~M., Tafalla, M., et al.\ 1999, \apjl, 523, L165

\bibitem[Caselli et al.(2002)]{Caselli2002} Caselli, P., Stantcheva, T., Shalabiea, O., et al.\ 2002, \planss, 50, 1257

\bibitem[Caselli et al.(2008)]{Caselli2008} Caselli, P., Vastel, C., Ceccarelli, C., et al.\ 2008, \aap, 492, 703

\bibitem[Ceccarelli et al.(2014)]{Ceccarelli2014} Ceccarelli, C., Caselli, P., Bockel{\'e}e-Morvan, D., et al.\ 2014, Protostars and Planets VI, 859




\bibitem[Feng et al.(2016)]{Feng2016} Feng, S., Beuther, H., Zhang, Q., et al.\ 2016, \aap, 592, A21


\bibitem[{{Fontani} {et~al.}(2012){Fontani}, {Giannetti}, {Beltr\'an}, {Dodson}, {Rioja},
{Brand}, {Caselli}, \& {Cesaroni}}]{Fontani2012}
{Fontani}, F., {Giannetti} A., {Beltr\'an} M.~T., {Dodson} R., {Rioja} M.,
{Brand} J., {Caselli} P., {Cesaroni} R. 2012, MNRAS, 423, 2342

\bibitem[Fontani et al.(2016)]{Fontani2016} Fontani, F., Commer{\c{c}}on, B., Giannetti, A., et al.\ 2016, \aap, 593, L14

\bibitem[Ford \& Shirley(2011)]{Ford2011} Ford, A.~B., \& Shirley, Y.~L.\ 2011, \apj, 728, 144


\bibitem[{{Giannetti} {et~al.}(2014){Giannetti}, {Wyrowski}, {Brand}, {Csengeri}, {Fontani}, {Walmsley}, {Nguyen Long}, {Beuther}, {Schuller}, {G\"usten}, \& {Menten}}]{Giannetti2014}
{Giannetti}, A., {Wyrowski}, F., {Brand}, J., {Csengeri}, T., {Fontani}, F., {Walmsley}, C. M., {Nguyen Luong}, Q., {Beuther}, H., {Schuller}, F., {G\"usten}, R., {Menten}, K. M.
, 2014, A\&A, 570, A65

\bibitem[Giannetti et al.(2017)]{Giannetti2017} Giannetti, A., Leurini, S., Wyrowski, F., et al.\ 2017, \aap, 603, A33


\bibitem[Giannetti et al.(2019)]{Giannetti2019} Giannetti, A., Bovino, S., Caselli, P., et al.\ 2019, \aap, 621, L7

\bibitem[Goodson et al.(2016)]{Goodson} Goodson, M.~D., Kong, S., Tan, J.~C., et al.\ 2016, \apj, 833, 274

\bibitem[{{Grassi} {et~al.}(2014){Grassi}, {Bovino}, {Schleicher}, {Prieto},
  {Seifried}, {Simoncini}, \& {Gianturco}}]{Grassi2014}
{Grassi}, T., {Bovino}, S., {Schleicher}, D.~R.~G., {et~al.} 2014, \mnras, 439,
  2386
%

\bibitem[Harju et al.(2006)]{Harju2006} Harju, J., Haikala, L.~K., Lehtinen, K., et al.\ 2006, \aap, 454, L55

\bibitem[Hasegawa et al.(1992)]{Hasegawa1992} Hasegawa, T.~I., Herbst, E., \& Leung, C.~M.\ 1992, \apjs, 82, 167

\bibitem[Hasegawa, \& Herbst(1993)]{Hasegawa1993} Hasegawa, T.~I., \& Herbst, E.\ 1993, \mnras, 261, 83

\bibitem[Hernandez et al.(2011)]{Hernandez2012} Hernandez, A.~K., Tan, J.~C., Caselli, P., et al.\ 2011, \apj, 738, 11


\bibitem[Hopkins(2015)]{Hopkins2015} Hopkins, P.~F.\ 2015, \mnras, 450, 53

\bibitem[Hopkins et al.(2018)]{Hopkins2017} Hopkins, P.~F., Wetzel, A., Kere{\v{s}}, D., et al.\ 2018, \mnras, 480, 800

\bibitem[Hugo et al. (2009)]{Hugo2009} Hugo, E., Asvany, O., Schlemmer, S. 2009, J. Chem. Phys., 130, 164302

\bibitem[Kalv{\={a}}ns(2016)]{Kalvans2016} Kalv{\={a}}ns, J.\ 2016, \apjs, 224, 42

\bibitem[K{\"o}nig et al.(2017)]{Koenig2017} K{\"o}nig, C., Urquhart, J.~S., Csengeri, T., et al.\ 2017, \aap, 599, A139


\bibitem[{{K\"ortgen} {et~al.}(2017){K\"ortgen}, {Bovino}, {Schleicher}, {Giannetti}, {Banerjee}}]{Koertgen2017} 
{K\"ortgen} B., {Bovino} S., {Schleicher} D.~R.~G., {Giannetti} A., {Banerjee} R.
2017, MNRAS, 469, 2602

\bibitem[{{K\"ortgen} {et~al.}(2018){K\"ortgen }, {Bovino}, {Schleicher}, {Stutz}, {Banerjee}, {Giannetti}, {Leurini}}]{Koertgen2018} {K\"ortgen} B., {Bovino} S., {Schleicher} D.~R.~G., {Stutz} A., {Banerjee} R., {Giannetti} A., {Leurini} S.2018, MNRAS, 478, 95

\bibitem[{{Kong} {et~al.}(2015){Kong}, {Caselli}, {Tan}, {Wakelam}, {Sipila}}]{Kong2015} 
{Kong}, S., {Caselli}, P., {Tan}, J. C., {Wakelam}, V., {Sipila}, O., 2015, ApJ, 804, 98

\bibitem[Leurini et al.(2011)]{Leurini2011} Leurini, S., Pillai, T., Stanke, T., et al.\ 2011, \aap, 533, A85


\bibitem[Lupi et al.(2018)]{Lupi2018} Lupi, A., Bovino, S., Capelo, P.~R., et al.\ 2018, \mnras, 474, 2884

\bibitem[Lupi et al.(2019)]{Lupi2019} Lupi, A., Volonteri, M., Decarli, R., et al.\ 2019, arXiv e-prints, arXiv:1901.02464

\bibitem[Mathis et al.(1977)]{Mathis1977} Mathis, J.~S., Rumpl, W., \& Nordsieck, K.~H.\ 1977, \apj, 217, 425


\bibitem[Motte et al.(2018)]{Motte2017} Motte, F., Bontemps, S., \& Louvet, F.\ 2018, \araa, 56, 41

\bibitem[Nguyen et al.(2018)]{Nguyen2018} Nguyen, T., Baouche, S., Congiu, E., et al.\ 2018, \aap, 619, A111

\bibitem[Padovani et al.(2009)]{Padovani2009} Padovani, M., Galli, D., \& Glassgold, A.~E.\ 2009, \aap, 501, 619

\bibitem[Pagani et al.(1992)]{Pagani1992} Pagani, L., Wannier, P.~G., Frerking, M.~A., et al.\ 1992, \aap, 258, 472

\bibitem[Pagani et al.(2009)]{Pagani2009} Pagani, L., Vastel, C., Hugo, E., et al.\ 2009, \aap, 494, 623

\bibitem[Pagani et al.(2012)]{Pagani2012} Pagani, L., Bourgoin A., and Lique F. \ 2012, \aa, 548, L4

\bibitem[Rathborne et al.(2006)]{Rathborne2006} Rathborne, J.~M., Jackson, J.~M., \& Simon, R.\ 2006, \apj, 641, 389

\bibitem[Redaelli et al.(2019)]{Redaelli2019} Redaelli, E., Bizzocchi, L., Caselli, P., et al.\ 2019, \aap, 629, A15

\bibitem[Sabatini et al.(2019)]{Sabatini2019} Sabatini, G., Giannetti, A., Bovino, S., et al.\ 2019, arXiv e-prints, arXiv:1910.02981

\bibitem[Sipil{\"a} et al.(2010)]{Sipila2010} Sipil{\"a}, O., Hugo, E., Harju, J., et al.\ 2010, \aap, 509, A98

\bibitem[Sipil{\"a} et al.(2015)]{Sipila2015} Sipil{\"a}, O., Caselli, P., \& Harju, J.\ 2015, \aap, 578, A55


\bibitem[Springel(2005)]{Springel2005} Springel, V.\ 2005, \mnras, 364, 1105

\bibitem[Troscompt et al.(2009)]{Trompscot2009} Troscompt, N., Faure, A., Maret, S., et al.\ 2009, \aap, 506, 1243


\bibitem[{{Truelove} {et~al.}(1997){Truelove}, {Klein}, {McKee}, {Holliman},
  {Howell}, \& {Greenough}}]{Truelove1997}
{Truelove}, J.~K., {Klein}, R.~I., {McKee}, C.~F., {et~al.} 1997, ApJL, 489,
  L179
  
\bibitem[Walmsley et al.(2004)]{Walmsley2004} Walmsley, C.~M., Flower, D.~R., \& Pineau des For{\^e}ts, G.\ 2004, \aap, 418, 1035

\bibitem[Wakelam et al.(2017)]{Wakelam2016} Wakelam, V., Loison, J.-C., Mereau, R., et al.\ 2017, Molecular Astrophysics, 6, 22

\bibitem[{{Zhang} {et~al.}(2009){Zhang}, {Wang}, {Pillai}, \& {Rathborne}}]{Zhang2009} 
{Zhang} Q., {Wang} Y., {Pillai} T., {Rathborne} J., 2009, ApJ, 696, 268



\end{thebibliography}

\end{document}